\documentclass{article}
\usepackage{spconf,amsmath,graphicx}
\usepackage{xcolor}
\usepackage[english]{babel}
\usepackage{bm}

\title{Speaker Mask Transformer for Multi-talker Overlapped Speech Recognition}
\name{Peng Shen, Xugang Lu, Hisashi Kawai\thanks{This work is partially supported by JSPS KAKENHI No.21K17776.}}
\address{National Institute of Information and Communications Technology (NICT), Japan}
\begin{document}
%\ninept
%
\maketitle
\begin{abstract}
  Multi-talker overlapped speech recognition remains a significant challenge, requiring not only speech recognition but also speaker diarization tasks to be addressed. In this paper, to better address these tasks, we first introduce speaker labels into an autoregressive transformer-based speech recognition model to support multi-speaker overlapped speech recognition. Then, to improve speaker diarization, we propose a novel speaker mask branch to detection the speech segments of individual speakers. With the proposed model, we can perform both speech recognition and speaker diarization tasks simultaneously using a single model. Experimental results on the LibriSpeech-based overlapped dataset demonstrate the effectiveness of the proposed method in both speech recognition and speaker diarization tasks, particularly enhancing the accuracy of speaker diarization in relatively complex multi-talker scenarios.

  %Multi-talker Ovalpped speech recognition 依然是ASR任务的一个很重要的挑战。本论文中，我们通过使用基于生成式的Transfomer模型，通过增加Mask branch来model speaker的语音区间。从而通过一个模型，可以同时来执行语音识别和Speaker diarization任务。结果显示，我们的模型可以更加准确的判定speaker的语音，并且可以减少SD的时间预测对ASR的影响。
\end{abstract}
\section{Introduction}
\label{sec:intro}

Recently, automatic speech recognition (ASR) techniques have achieved significant progress, thanks to advancements in system architecture and optimization algorithms \cite{jinyuLi2021E2EASR,Chan2016ListenAA,Li2020DevelopingRM,Yu2014ASR}. However, real-world application scenarios present significant challenges, notably a decrease in recognition accuracy rate, for example the multi-talker speech transcription tasks \cite{YuDong2017,KandaSOT2020}. In such tasks, the major challenge is the influence of overlapped speech. Systems must not only recognize overlapped speech but also detect individual speakers' speech segments to further utilize speaker information for enhancing ASR or facilitating human verification.

Multi-talker overlapped speech recognition tasks have been investigated by many researchers \cite{YuDong2017,KandaSOT2020,chang2020,NTToverlap2023,KandaStreaming2022}. A common solution is to undertake a two-step process that initially separates overlapped speeches into individual speeches using techniques such as deep clustering \cite{Isik2016, Hershey2016}. Subsequently, a typical ASR system, optimized for single-speaker speech, is employed to transcribe the separated speeches into text \cite{chang2020, Chorowski2015}.
Recently, autoregressive model-based approaches have also been proposed \cite{KandaSOT2020, KandaStreaming2022}. These approaches directly generate the transcription of multiple speakers recursively, one after another. Unlike the two-stage approaches, these methods use the same architecture as single-talker ASR systems, and use designing labels with speaker changes and text-based tokens to predict speaker changes and ASR \cite{KandaStreaming2022}.

To further predict timestamp information with the ASR model, previous works have utilized an autoregressive model to predict the start and end times of an utterance by using quantized timestamp tokens \cite{whisper2022}. This idea has also been explored in the context of multi-talker overlapped speech recognition tasks \cite{NTToverlap2023}. In their work, they introduce special tokens to present the start and end timestamps of utterances in the labels, which are presented alongside speaker change labels during model training.
However, when considering the complexity of real meeting discussion speech data with frequent speaker changes (dialogues are not simply one speaker following another, and interruptions often occur), the labels need include the timestamp special tokens for all the speaker changes. We argue that these artificially defined special tokens are not components of natural language and may potentially distort the language model's representation of the linguistic context.

To overcome this issue, we first introduce speaker labels for distinguish different speakers. Then, inspired by Mask R-CNN \cite{MaskRCNN2018}, we propose a speaker mask branch for detecting the target speaker's speech segments. With this design, our model can not only perform speech recognition but also identify different speakers and their corresponding speech segments. To the best of our knowledge, this approach has not been verified in multi-talker overlapped ASR tasks.
The contributions of this work can be summarized as follows: (1) We propose a novel speaker mask branch coupled with speaker labels for the transformer model, aiming to perform multi-talker overlapped speech recognition and speaker diarization. (2) Several configurations of the speaker mask branch are designed and evaluated. (3) We prepare a simple and a relatively complex dataset to verify the performance of the proposed method, and further investigate its performance with a large-scale model.

%\vfill
%\pagebreak

\section{Multi-talker Overlapped Speech Recognition}
\label{sec:whisper}
\vspace{-0.5em}
% 在multi-speaker的ASR任务中，比如meeting transcritpion，通常我们的任务有两个，第二，对所有speaker的语音做ASR识别，第二，检测出各个speaker的的语音区间。也就是通常说的，what，when，what。

\begin{figure*}
  \centering
  \includegraphics[width=440pt]{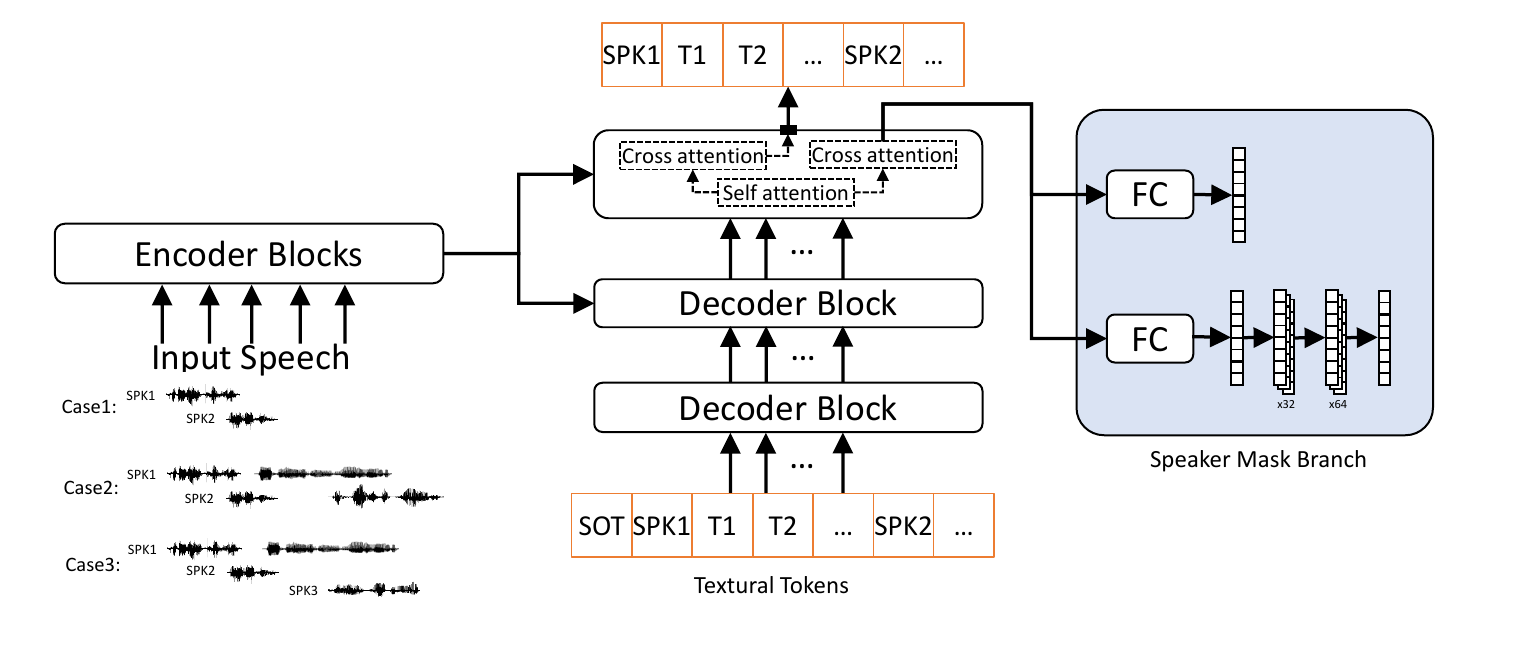}\\ \relax
  \caption{Overview of the proposed methods: (FC refers to the fully-connected layer, SOT is the Start of Transcription, SPK1 and SPK2 refer to speakers,  T1 and T2 are tokens related to text.)}
  \label{fig.propose_method1}
\end{figure*}
\vspace{-0.5em}

\subsection{Task definition}
\label{sec:taskdefinition}
In this work, we focus on developing multi-talker ASR techniques for meeting transcription tasks, therefore, we have two primary objectives. First, we need to detect WHO is speaking and the speech segments of each speaker to determine WHEN they are speaking. Secondly, we need to perform ASR on the speech of all speakers to transcribe WHAT they are saying.

One of the main challenges of this task is handling overlapped speech. In real scenarios, the situations where speech overlaps can be complex. For instance, as illustrated in Fig. \ref{fig.propose_method1}, overlaps can occur not only with two speakers, as seen in Case 1, but also in situations where two speakers take turns speaking. Even more complex are scenarios involving the participation of multiple speakers, like in Case 3. In this paper, our experiments primarily focus on the circumstances presented in Case 1 and Case 2.

\vspace{-0.5em}
\subsection{Label-based speaker prediction and diarization}
\label{sec:label_based}
Previous works have proposed the use of generative models for ASR and timestamp prediction \cite{NTToverlap2023, whisper2022}. In studies such as \cite{KandaSOT2020, NTToverlap2023, KandaStreaming2022}, a speaker change symbol is employed to denote transitions between speakers. While this single special token cannot track all speech segments attributed to a single speaker throughout the entire utterance. In our task, to track the speakers more effectively, we introduce several speaker labels in place of the speaker change symbol.

Defined $\mathbf{X} $ and $\mathbf{W}$ as the acoustic feature and textual tokens of a speech utterance, respectivelly. We assume that the speech utterances was spoken by $K$ speakers, denoted as $S_{1:K}$, and their corresponding textual tokens are denoted as $\mathbf{W}_{1:K}$. The start and end times for $k$-th speaker in a utterance are defined as $T^k_s$ and $T^k_e$, respectively.
Similar to the approach adopted in the Whisper model, $T^k_s$ and $T^k_e$ are quantized time tokens obtained by rounding the continuous timestamp values to the nearest quantized value at intervals of every 20ms \cite{whisper2022}.
Following the timestamp settings of the Whisper model, we can define the labels as follows:

\begin{equation}\label{eq:method.01}
  \begin{split}
    \mathbf{U}_{\text{SPK-TS-1}} = <S_1> <T^1_s> \mathbf{W}_1 <T^1_e> \dots \\ <S_k> <T^k_s> \mathbf{W}_k <T^k_e> \dots
  \end{split}
\end{equation}
where $<\bullet>$ denotes special tokens that differ from the tokens used in original speech recognition. Similar to previous works \cite{KandaSOT2020,NTToverlap2023}, the order of the speaker labels follows a first-in, first-out principle, meaning that the results from the individual who spoke first will be output first. Referring to the timestamp setting outlined in \cite{NTToverlap2023}, we can also define the labels as follows:
\begin{equation}\label{eq:method.02}
  \begin{split}
    \mathbf{U}_{\text{SPK-TS-2}} = <S_1> <T^1_s> <T^1_e> \mathbf{W}_1 \dots \\ <S_k> <T^k_s> <T^k_e> \mathbf{W}_k \dots
  \end{split}
\end{equation}
With the defined labels, the model can be optimized to handle speaker detection, timestamp prediction, and ASR, using the following objective functions:
\begin{equation}\label{eq:strategy2.03}
  \begin{split}
    L_{\text{ASR}} = -\log P(\mathbf{U}  | \mathbf{X} , \mathbf{\Theta} )
  \end{split}
\end{equation}
where $\mathbf{U}$ can use the definition of Eq.\ref{eq:method.01} and Eq.\ref{eq:method.02}, and $\mathbf{\Theta}$ denotes the model parameters.

\section{Proposed Speaker Mask Method}
\label{sec:propsoed}

The label-based methods predict the speaker's speech segments by adding speaker labels and start and end timestamps to the labels. However, as we discussed in section \ref{sec:taskdefinition}, in actual meetings, situations of voice overlap are often more complex. Solely using start and end timestamps cannot effectively detect the intervals in Case 2 (Fig. \ref{fig.propose_method1}). Motivated by Mask R-CNN \cite{MaskRCNN2018}, we introduce a new speaker mask branch to the transformer model to improve speaker diarization.

Formally, for a $D$-dimensional hidden representation of the speech signal (encoder outputs), the mask branch yields an output that maintains the same $D$ dimensions.
%The speaker mask label, denoted as $\mathbf{Y}$, is a $D$-dimensional binary vector; a value of 0 indicates non-target speaker speech, while a value of 1 denotes target speaker speech. 
For each output dimension, we apply a per-dimensional sigmoid function and use the average binary cross-entropy loss to optimize the model that is described as follows:
\begin{equation}\label{eq:proposed.01}
  \begin{split}
    L^k_{\text{mask}} = -\frac{1}{D}\sum_{i=1}^{D} \left[ y_i \log(p_i) + (1 - y_i) \log(1 - p_i) \right]
  \end{split}
\end{equation}
where \( y_i \) represents the true labels (which will be 0 or 1), and \( p_i \) represents the predicted probabilities.
%The loss is calcuated over all the output and take the average over these to get the average loss.

During training, the mask branch is optimized only when the main branch outputs speaker labels; therefore, the multi-task objective function can be defined as,
\begin{equation}\label{eq:proposed.02}
  \begin{split}
    L_{\text{SPK-MASK}} = (1-\lambda) * L_{ASR} + \lambda \sum_{k=1}^{K} L^k_{mask}
  \end{split}
\end{equation}
where $\lambda$ is the weight coefficient used to balance the original ASR output loss and the speaker mask loss. For the labels of ASR main branch, we remove the timestamps in Eq.\ref{eq:method.01}, which can be rewritten as follows:
\begin{equation}\label{eq:proposed.03}
  \begin{split}
    \mathbf{U}_{\text{SPK}} = <S_1> \mathbf{W}_1 \dots <S_k> \mathbf{W}_k \dots
  \end{split}
\end{equation}

\section{Experiment and Results}
\label{sec:prior}

\subsection{Dataset and evaluation metrics}
To evaluate the proposed method, we utilized the LibriSpeech corpus to simulate multi-talker overlapped speech data.
For training dataset, all the 960 hours of training data was utilized.
We prepared two types of training datasets for Case 1 (two-speaker simple case) and Case 2 (two-speaker complex case).
For Case 1, we randomly selected an utterance from a different speaker for each utterance and mixed them at a signal-to-interference ratio of 0 dB.
In Case 2, we chose two utterances from speaker 1 and one utterance from another speaker, speaker 2, mixing them in the order of Speaker 1, Speaker 2, Speaker 1. The overlap duration between every two utterances was also randomly determined, ranging from 0 to 5 seconds.
Finally, aside from the original dataset, i.e., "train960-Org", we obtained "train960-Set1" which combined the original and Case 1 training data at a 1:1 ratio, and "train960-Set2" that included the original, Case 1, and Case 2 training data at a 1:1:1 ratio.

The evaluation set was generated base on "test-clean", in the same way to the training data, the only difference being that we fixed the overlap duration to 1s and 3s. This resulted in three evaluation sets: "Set1-1s" and "Set1-3s" for Case 1, and "Set2-1s" for Case 2.
We used the time-alignment generated by the Montreal Forced Aligner \cite{ForcedAligner2017} to establish the start and end timestamps for each utterance. For labels for the speaker mask branch, we employed an energy-based VAD to discern speech or non-speech intervals in 20ms segments. Time-alignment timestamps were utilized to fix the start of the frames, reducing mis-determinations by the VAD due to noise.

For the speaker diarization evaluation, we applied diarization error rate (DER) with collar parmater of 0.2 and speaker count accuracy (SCA) as evaluation metrics. For ASR, we used the word error rate (WER) as the metric.

\begin{table*}[tb]
  \centering
  \caption{Experimental results (\%) of baseline and proposed methods with "train960-Set1" datasets for Case 1.}
  \setlength{\tabcolsep}{0.4em}
  \begin{tabular}{|l||c||c|c|c||c|c|c|c|c|c|c|c|c|c|c|} \hline
                                                  & test-clean    & \multicolumn{3}{c||}{Set1-1s} & \multicolumn{3}{c|}{Set1-3s}                                                 \\ \hline
    Methods ($\lambda$)                           & WER           & WER                           & DER                          & SCA   & WER           & DER           & SCA   \\ \hline \hline
    Pyannote.audio \cite{Bredin2020}              & -             & -                             & 10.08                        & -     & -             & 14.07         & -     \\ \hline
    Single-talker:Whisper base                    & 5.87          & 25.66                         & -                            & -     & 48.20         & -             & -     \\ \hline
    Single-talker:Whisper base FT on train960-Org & \textbf{3.46} & 11.88                         & -                            & -     & 29.70         & -             & -     \\ \hline

    Multi-talker: SPK                             & 3.59          & 5.93                          & -                            & 99.50 & \textbf{9.47} & -             & 99.09 \\ \hline
    Multi-talker: SPK-TS-1                        & 4.06          & 8.14                          & 2.72                         & 99.89 & 13.13         & 3.09          & 99.90 \\ \hline
    Multi-talker: SPK-TS-2                        & 3.75          & 6.25                          & 2.92                         & 99.73 & 10.74         & 3.33          & 99.45 \\ \hline\hline

    Mask: L-FC (0.1)                              & 3.65          & 5.98                          & 3.55                         & 99.20 & 9.69          & 4.16          & 98.59 \\ \hline
    Mask: L-FC (0.5)                              & 3.72          & 6.05                          & 1.99                         & 99.54 & 10.10         & 2.13          & 99.60 \\ \hline
    Mask: L-FC (0.9)                              & 3.83          & 6.62                          & \textbf{1.14}                & 99.62 & 10.92         & \textbf{1.63} & 99.09 \\ \hline\hline
    Mask: L-FC-CNN (0.5)                          & 3.60          & \textbf{5.88}                 & 1.87                         & 99.62 & 9.97          & 2.15          & 99.25 \\ \hline
    Mask: CA-FC-CNN (0.5)                         & 3.70          & 6.03                          & \textbf{1.54}                & 99.69 & 9.91          & \textbf{2.04} & 99.65 \\ \hline
  \end{tabular}
  \vspace{-3mm}
  \label{result.case1}
\end{table*}

\begin{table*}[tb]
  \centering
  \caption{Experimental results (\%) of baseline and proposed methods with "train960-Set2" training datasets for Case 2.}
  \setlength{\tabcolsep}{0.3em}
  \begin{tabular}{|l||c||c|c|c||c|c|c||c|c|c|c|c|c|c|c|} \hline
                                   & test-clean    & \multicolumn{3}{c||}{Set1-1s} & \multicolumn{3}{c||}{Set1-3s} & \multicolumn{3}{c|}{Set2-1s}                                                                                  \\ \hline
    Methods  ($\lambda$)           & WER           & WER                           & DER                           & SCA                          & WER           & DER           & SCA    & WER           & DER           & SCA   \\ \hline \hline

    Multi-talker: SPK              & 3.98          & \textbf{5.73}                 & -                             & 100.00                       & \textbf{8.86} & -             & 99.90  & \textbf{6.51} & -             & 99.85 \\ \hline
    Multi-talker: SPK-TS-2         & 4.15          & 5.98                          & 2.71                          & 99.73                        & 9.26          & 2.91          & 99.60  & 6.69          & 26.37         & 99.76 \\ \hline\hline
    Mask: L-FC (0.5)               & \textbf{3.86} & 5.86                          & 0.71                          & 99.69                        & 8.89          & 1.14          & 99.90  & 6.59          & 1.10          & 99.95 \\ \hline
    Mask: L-FC-CNN (0.5)           & 4.04          & 5.88                          & 0.77                          & 99.73                        & 9.18          & 1.17          & 99.75  & 6.69          & 1.01          & 99.80 \\ \hline
    Mask: CA-FC-CNN (0.5)          & 3.98          & 5.98                          & \textbf{0.60}                 & 99.77                        & 8.91          & \textbf{1.02} & 100.00 & \textbf{6.51} & \textbf{0.70} & 99.85 \\ \hline\hline

    LargeV2,Multi-talker: SPK      & \textbf{1.84} & 2.49                          & -                             & 99.81                        & 4.43          & -             & 99.80  & \textbf{2.53} & -             & 99.80 \\ \hline
    LargeV2,Multi-talker: SPK-TS-2 & 2.56          & 2.90                          & 2.54                          & 99.69                        & 4.63          & 2.72          & 99.75  & 2.88          & 26.25         & 99.75 \\ \hline
    LargeV2,Mask: L-FC (0.5)       & 1.88          & \textbf{2.46}                 & \textbf{0.76}                 & 99.96                        & \textbf{4.33} & \textbf{1.10} & 99.90  & 2.69          & \textbf{0.80} & 99.90 \\ \hline
  \end{tabular}
  \vspace{-3mm}
  \label{result.case2}
\end{table*}

\vspace{-0.5em}
\subsection{Experimental setup}
\vspace{-0.5em}

To compare with the proposed method, we prepared several baseline systems. One is based on the pyannote.audio toolkit \cite{Bredin2020}, utilizing the released speaker diarization model for speaker diarization \footnote{https://huggingface.co/pyannote/speaker-diarization}. The other two baseline methods, referred to as the label-based methods, are described in section \ref{sec:label_based} — specifically, the methods corresponding to Eqs. \ref{eq:method.01} ("Multi-talker: SPK-TS-1") and \ref{eq:method.02} ("Multi-talker: SPK-TS-2"). For the proposed method, we employ Eq. \ref{eq:proposed.03} as labels for the main outputs, and the model was optimized using Eq. \ref{eq:proposed.02}. In all three methods, the model was based on the Whisper base model \cite{whisper2022}, which comprises 74M parameters, features 6 transformer blocks in both the encoder and decoder, each with a hidden dimension of 512. The input features utilized are 80-channel log-magnitude Mel spectrograms computed on 25ms windows with a stride of 10ms.

For the speaker mask branch, we evaluated two settings. In the first setting, we directly used a fully connected layer to convert the output of a new cross-attention block into the final mask, utilizing a 1500 dimension configuration ("Mask:CA-FC"). In the second approach ("Mask:CA-FC-CNN"), we employ two CNN layers with 32 and 64 kernels, respectively, each having a kernel size of 2x1, a stride of 1, and appropriate padding settings. A dropout layer with a rate of 0.25 is applied after the second CNN layer. Apart from using the new cross-attention block, we also experiment with using the last output logits as inputs to the speaker mask branch; these two strategies are referred to as "Mask:L-FC" and "Mask:L-FC-CNN", respectively.

We trained the model using the Adam algorithm accompanied by a warm restart learning rate scheduler during the training process \cite{Loshchilov2016Warmup}. The global mini-batch size was set to 192 with an initial learning rate of 1e-4, and a minimum learning rate of 1e-8, adjusted according to a cosine annealing-based learning rate schedule. The number of epochs was set to 10.

%In our proposed method, we carried out investigations utilizing both the Whisper base and large-v2 models \cite{whisper2022}. Both these models are based on transformer networks. The base model comprises 74M parameters, featuring 6 transformer blocks in both the encoder and decoder, each with a hidden dimension of 512. The large-v2 model contains 1550M parameters and includes 32x2 transformer blocks with a hidden dimension of 1280, making it a significantly large model for speech processing tasks. The input features used are 80-channel log-magnitude Mel spectrograms computed on 25ms windows with a stride of 10ms. The language embedding block is same to the ETDNN baseline. We opted not to use the SpecAugment technique in the proposed method.

\vspace{-0.5em}
\subsection{Investigation and experimental results}
\vspace{-0.5em}

In Table \ref{result.case1}, the experimental results for Case 1 involving several baselines and the proposed method are listed. From the results, we observe that the single-talker model, even after fine-tuning on LibriSpeech (denoted as "train960-Org"), struggles to perform well on the overlapped speech dataset. However, by incorporating speaker labels — as in "Multi-talker: SPK" — the performance on the overlapped dataset significantly improves. When comparing with the pyannote.audio models, both "SPK-TS-1" and "SPK-TS-2" managed to reduce the DER, but with a WER increase across all three datasets.
For the proposed speaker mask method, the "Mask: L-FC" approaches, especially with a $\lambda$ value of 0.9, obtained the lowest DER, but with a large performance degradation for ASR tasks.
The results indicate that setting $\lambda$ to 0.5 provides a better balance between WER and DER.
With $\lambda$ set to 0.5, we further investigated the branch networks by adding CNN layers ("Mask: L-FC-CNN") and employing a new cross-attention setting ("Mask: CA-FC-CNN"), witnessing consistent improvement in DER with no change or even a reduction in WER.

For Case 2 (results in Table \ref{result.case2}), we observed trends almost same to those of Case 1. Comparatively, the "Multi-talker: SPK-TS-2" did not perform well for Case 2 (i.e., results of Set2-1s), with a DER of 26.37\%. The proposed method achieved a DER almost comparable to that in Case 1. When compared to models trained with "train960-Set1", those trained with "train960-Set2" exhibited lower WER and DER on overlapped datasets, albeit with a slightly higher WER on the test-clean dataset.

To further examine the influence of model parameters, we conducted investigations based on the Whisper Large-v2 model, training models with a learning rate of 1e-5 and setting the number of epochs to 5. Although we observed a significant reduction in WER across all datasets, the DER did not see substantial degradation. This can be primarily attributed to the new tasks not being tuned during the pre-training of the original Whisper model.

%Overall, the proposed speaker mask method significantly reduced the DER for speaker diarization, both in the simpler Case 1 and the relatively more complex Case 2, with only a minimal increase in WER.

\vspace{-0.5em}
\section{Conclusions}
\vspace{-0.5em}
In this study, we proposed a speaker mask method for speaker diarization in transformer-based multi-talker speech recognition systems. Utilizing the speaker mask branch, we demonstrated that our model could reduce DER effectively in both simple and relatively complex scenarios. When compared to methods that add timestamps to labels used for ASR training, the proposed method also exerts a lower influence on the accuracy of ASR.

\vfill\pagebreak

%\section{REFERENCES}
%\label{sec:refs}

\bibliographystyle{IEEEbib}
\bibliography{mybib4asr, refs}

\end{document}